# Secure Communication and Access Control for Mobile Web Service Provisioning

Satish Narayana Srirama[1], Anton Naumenko[2]

[1]RWTH Aachen University, Informatik V (Information Systems)
Ahornstr 55, 52056 Aachen, Germany
[2]Industrial Ontologies Group, Department of Mathematical Information Technology,
P.O. Box 35, 40014 University of Jyväskylä, Finland
srirama@cs.rwth-aachen.de, annaumen@cc.jyu.fi

**Abstract.** It is now feasible to host basic web services on a smart phone due to the advances in wireless devices and mobile communication technologies. While the applications are quite welcoming, the ability to provide secure and reliable communication in the vulnerable and volatile mobile ad-hoc topologies is vastly becoming necessary. The paper mainly addresses the details and issues in providing secured communication and access control for the mobile web service provisioning domain. While the basic message-level security can be provided, providing proper access control mechanisms for the Mobile Host still poses a great challenge. This paper discusses details of secure communication and proposes the distributed semantics-based authorization mechanism.

**Keywords:** Access Control, Communication system security, Mobile Communication, Mobile web services.

## 1 Introduction

The high-end mobile phones and PDAs are becoming pervasive and are being used in variety of applications like location based services, banking services, ubiquitous computing etc. The higher data transmission rates achieved with 3G and 4G technologies also boosted this growth in the wireless market. The situation brings out a large scope and demand for software applications for such high-end mobile devices. To meet this demand and to reap the benefits of the fast growing web services domain and standards, the scope of the mobile terminals as both web services clients and providers is being observed. While mobile web service clients are common these days, we have studied the scope of mobile web service provisioning, in one of our previous projects. [5]

Mobile web service provisioning offers many of its applications in domains like collaborative learning, social systems, mobile community support etc. While the applications are quite welcoming, the ability to provide secure and reliable communication in the vulnerable and volatile mobile ad-hoc topologies is vastly becoming necessary. Moreover with the easily readable mobile web services, the complexity to realize security increases further. Secure provisioning of mobile web services needs proper identification mechanism, access control, data integrity and confidentiality.

In our current research, we are trying to provide proper security for the mobile web service provider ("Mobile Host") realized by us. The security analysis suggests that proper message-level security can be provided in mobile web service provisioning with reasonable performance penalties on the Mobile Host. While the basic message-level security can be provided, the end-point security comprising proper identity and access control mechanisms, still poses a great challenge for the Mobile Host. Here we propose to utilize distributed architectures of semantics-based authorization mechanism to ensure pro-active context-aware access control to mobile web services.

The rest of the paper is organized as follows: Section 2 discusses the concept and analysis of mobile web service provisioning domain. Section 3 addresses the issues of securing the communication for mobile web services. Section 4 presents our research ideas towards implementation of semantics-based access control for mobile web services and section 5 concludes the paper with future research directions.

## 2 Pervasive Mobile Web Service Provisioning

Traditionally, the hand-held cellular devices have many resource limitations like limited storage capacities, low computational capacities, and small display screens with poor rendering potential. Most recently, the capabilities of these wireless devices like smart phones, PDAs are expanding quite fast. This is resulting in quick adoption of these devices in domains like mobile banking, location based services, social networks, e-learning etc. The situation also brings out a large scope and demand for software applications for such high-end wireless devices.

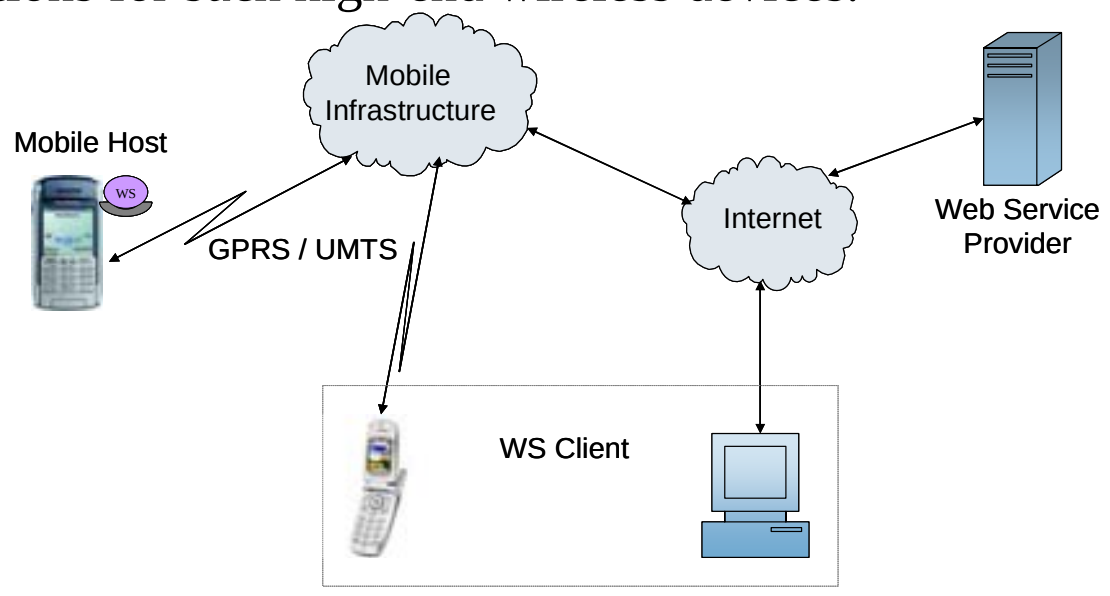

**Fig. 1.** Mobile terminals as both web service providers and clients

Moreover, with the achieved high data transmission rates in cellular domain, with interim and third generation mobile communication technologies like GPRS, EDGE and UMTS [2], mobile phones are also being used as Web Service clients and providers, bridging the gap between the wireless networks and the stationery IP networks. Combining these two domains brings us a new trend and lead to manifold opportunities to mobile operators, wireless equipment vendors, third-party application developers, and end users [4, 5]. While mobile web service clients are quite common these days, and many development tools are available from major vendors, the research with mobile web service provisioning is still sparse [3, 4]. During one of our previous projects, a small mobile web service provider ("Mobile Host") has been developed for resource constrained smart phones. Figure 1 shows the scenario with mobile terminal as both web service provider and client.

Mobile Host is a light weight web service provider built for resource constrained devices like cellular phones. It has been developed as a web service handler built on top of a normal web server. The web service requests sent by HTTP tunneling are diverted and handled by the web service handler. The Mobile Host was developed in PersonalJava [7] on a SonyEricsson P800 smart phone. The footprint of our fully functional prototype is only 130 KB. Open source kSOAP2 [8] was used for creating and handling the SOAP messages.

The detailed evaluation of the Mobile Host clearly showed that service delivery as well as service administration can be performed with reasonable ergonomic quality by normal mobile phone users. As the most important result, it turns out that the total WS processing time at the Mobile Host is only a small fraction of the total request-response invocation cycle time (<10%) and rest all being transmission delay [5]. This makes the performance of the Mobile Host directly proportional to achievable higher data transmission rates. Thus the high data transmission rates achievable, in the order of few Mbps, with advanced mobile communication technologies in 2.5G, 3G and 4G, help in realizing these Mobile Hosts also in commercial applications [9].

Mobile Host opens up a new set of applications and it finds its use in many domains like mobile community support, collaborative learning, social systems and etc. Many applications were developed and demonstrated, for example in a distress call; the mobile terminal could provide a geographical description of its location (as pictures) along with location details. Another interesting application scenario involves the smooth co-ordination between journalists and their respective organizations. From a commercial viewpoint, Mobile Host also renders possibility for small mobile operators to set up their own mobile web service businesses without resorting to stationary office structures. [5]

The Mobile Hosts in an operator proprietary network can also form a P2P network with other mobile phones and can share their individual resources and services. P2P offers a large scope for many applications with Mobile Host. Not just the enhanced application scope, the P2P network also offers better identification and discovery mechanisms of huge number of web services possible with Mobile Hosts. [10]

While the applications possible with Mobile Host are quite welcoming in different domains, the ability to provide the secured communication and access control in the vulnerable and volatile mobile ad-hoc topologies is quite challenging.

## 3 Securing Message Communication of Mobile Web Services

Once a mobile web service is developed and deployed with the Mobile Host, the service and the provider are prone to different types of security breaches like denial-of-service attacks, man-in-the-middle attacks, intrusion, spoofing, tampering etc. As web services use message-based technologies for complex transactions across multiple domains, traditional point-to-point security paradigms fall short. Potentially, a web-service message traverses through several legitimate intermediaries before it reaches its final destination. The intermediaries can read, alter or process the message. Therefore, the need for sophisticated end-to-end message-level security becomes a high priority and is not addressed by existing security technologies and standards in the wireless domain.

At the minimum, the mobile web service communication should possess the basic security requirements like proper authentication/authorization, and confidentiality/Integrity. Secure message transmission is achieved by ensuring message confidentiality and data integrity, while authentication and authorization will ensure that the service is accessed only by the legitimate service requestors. Even though a lot of security specifications, protocols like WS-Security [11], SAML [12] etc., exist for web services in traditional wired networks, not much has been explored and standardized in wireless environments. Our study contributes to this work and tries to bridge this gap, with main focus at realizing some of the existing security standards in the mobile web services domain.

The WS-Security specification from OASIS is the core element in web service security realm in wired networks. It provides ways to add security headers and tokens, insert timestamps, and to sign and encrypt the SOAP messages. To adapt the WS-Security in the mobile web service communication, the web service messages were processed with different encryption algorithms, signer algorithms and authentication principles, and were exchanged according to the standard. The performance of the Mobile Host was observed during this analysis, for reasonable quality of service. The main parameters of interest were the extra delay and variation in stability of the Mobile Host with the launched security overhead.

For the analysis, a SonyEricsson P910i smart phone was used. The device supports J2ME MIDP2.0 [13] with CLDC1.0 [14] configuration. For cryptographic algorithms and digital signers, java based light weight cryptographic API from Bouncy Castle crypto package [15] was used. kSOAP2 was modified and adapted according to WS-Security standard and utilized to create the request/response web service messages. The Mobile Host was redesigned with J2ME and the adapted kSOAP2.

To achieve confidentiality, the web service messages were ciphered with symmetric encryption algorithms and the generated symmetric keys were exchanged by means of asymmetric encryption methods. The messages were tested against various symmetric encryption algorithms, along with the WS-Security mandatory algorithms, TRIPLEDES, AES-128, AES-192 and AES-256 [16]. The PKI algorithm used for key exchange was RSA-V1.5 with 1024 and 2048 bit keys. Upon successful deployment of confidentiality, we considered data integrity on top of confidentiality. The messages were digitally signed and were evaluated against two signature algorithms, DSAwithSHA1 (DSS) and RSAwithSHA1. [6]

Figure 2 shows the total times taken at the Mobile Host for processing the web service requests. The timestamps consider the added security overload, for different symmetric algorithms on Mobile Host, but exclude the transmission delays. The test configuration considered here was RSA-1024 key exchange and RSA signature. This test is conducted against varied soap message sizes ranging from 1 to 10 KB. The extra delay for highly secured communication, AES-256 bit ciphered was approximately ~3 sec with RSAwithSHA1 signature for reasonable message sizes of 5KB. The delay was reasonable with respect to the Mobile Host's processing capability. [5]

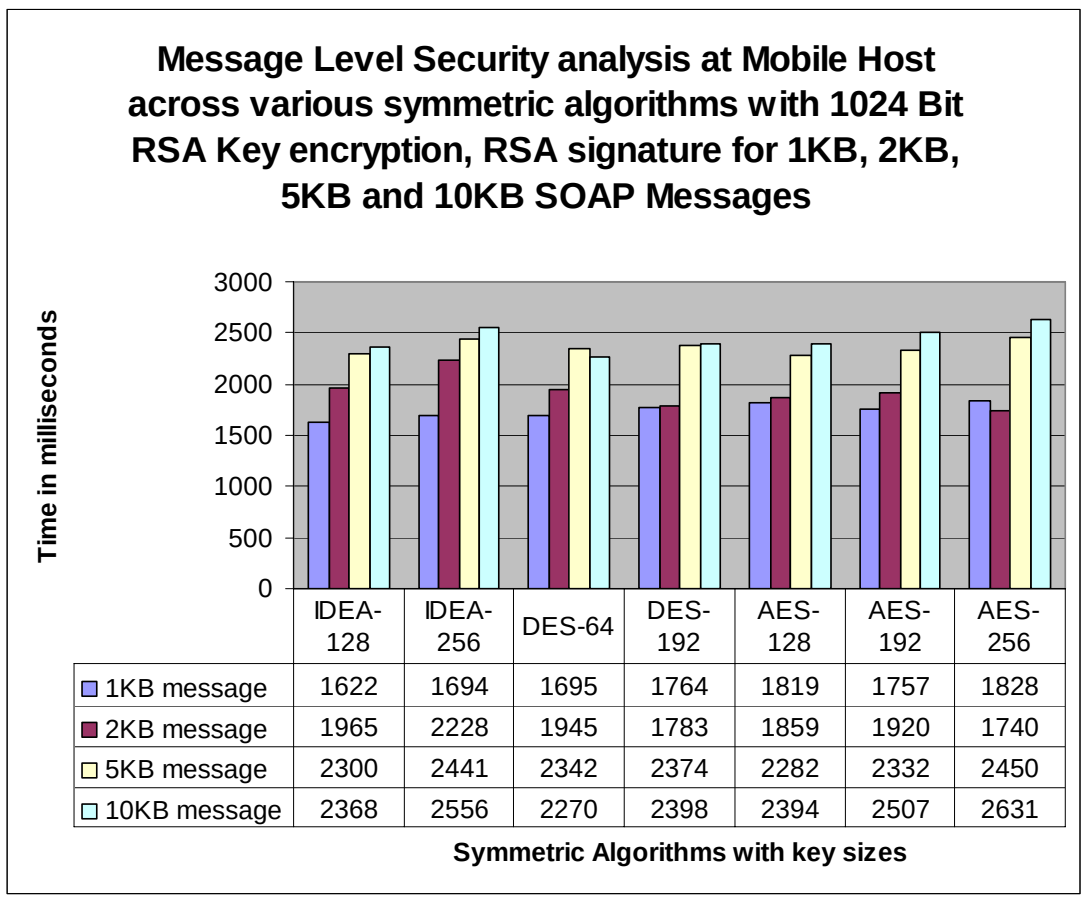

**Fig. 2.** Mobile Host's processing times for various message sizes and symmetric key algorithms using RSA signature

The detailed performance analysis suggested that not all of the WS-Security specification can be adapted to the mobile web service communication with today's smart phones. But with latest developments in speed and performances of processor chips for mobile devices, the scenario is going to change soon. With our security study for today's smart phones, we are recommending that the best way of securing SOAP messages in mobile web service provisioning is to use AES symmetric encryption with 256 bit key for encrypting the message and RSAwithSHA1 to sign the message. The symmetric keys are to be exchanged using RSA 1024 bit asymmetric key exchange mechanism. The cipher data and the keys are to be incorporated into the SOAP message according to WS-Security specification.

## 4 Semantics-based Access Control Mechanisms

For the trusted and distributed management of access control to protect mobile web services, we propose to use Semantics-Based Access Control (SBAC). SBAC is the result of adoption of the Semantic Web vision and standards [22] to the access control research and development field. Administration and enforcement of access control policies based on semantics of web services, clients, mediating actors and domain concepts comprise the most suitable approach to handle openness, dynamics, mobility, heterogeneity, distributed nature of environments that are involved in the mobile web service provisioning. We have previously defined the SBAC research framework [17], the SBAC model [18] and ontologies [19], the SBAC abstract architecture [17-21], conducted quantitative evaluation of the prototype of the SBAC policy enforcement function [19], and described adoption of SBAC for Semantic Web Services [20] and Multi-Agent Systems [21]. This paper covers the adoption of SBAC for the mobile SOA providing the analytical feasibility study of SBAC deployment options. This analytical evaluation takes into account results of experiments with the prototype of the SBAC enforcement function [19]. Before the analysis, we give a short description of the SBAC research framework, model, ontologies, and abstract architecture.

The SBAC research framework defines the SBAC research and development outcomes and their interrelations [17]. The whole framework reuses achievements in the Semantic Web research and development area like standards, ontologies, methodologies, frameworks, tools, platforms, etc. For example, the model-theoretic semantics of SBAC [18] is an extension of the direct model-theoretic semantics defined in the Web Ontology Language (OWL) standard [23] and Semantic Web Rule Language (SWRL) [24]. The SBAC model has been expressed in the form of ontologies [19]. Thus ontologies are the key part of the SBAC framework [17]. The ontology engineering constitutes the traditional domain modeling.

The abstract SBAC architecture is the upper view on components of SBAC and interactions between them to provide authorized access to protected resources. Basically, the abstract architecture is a bridge between theoretical findings and adoption of the SBAC into practice. Figure 3 shows elements of the SBAC abstract architecture encompassing the SBAC enforcement function only.

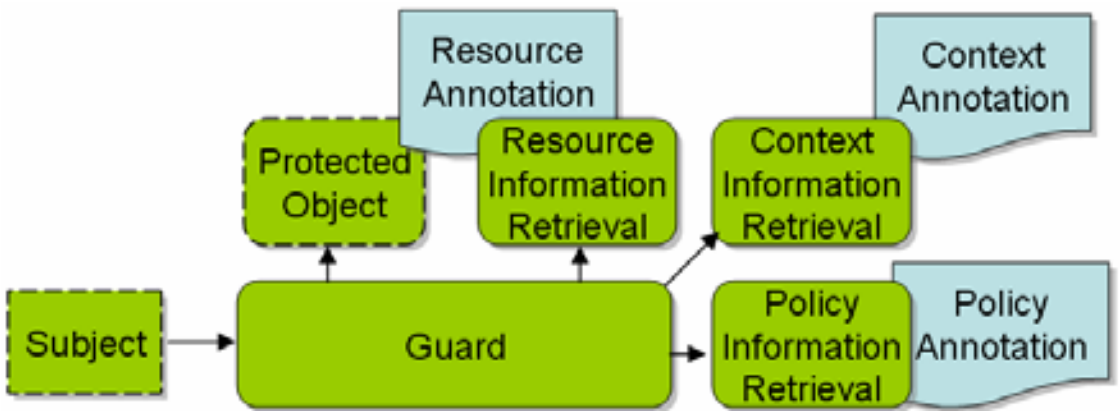

**Fig. 3.** The SBAC abstract architecture for the enforcement function.

The elements of the SBAC enforcement mechanism are the pro-active guard, the unified information retrieval components for policies (PIR), contextual data (CIR), and resource annotations (RIR). The guard is a proxy for both protected resources and information retrieval components. The guard pro-actively collects all relevant data and enforces an access decision based on the iterative reasoning over the semantically encoded access control policies and the semantic annotations of a subject, an operation, an object and a context of access [18, 20].

For the authorized mobile web service provisioning with the help of Mobile Host, the meanings of some crucial generic elements of the SBAC architecture are:

-- A subject of access is a human user who invokes the protected web services on the Mobile Host using a mobile phone or regular computer through Internet and mobile networks.

-- An operation of access is a web service's operation itself.

-- An object of access represents a protected web service deployed on the Mobile Host together with the input values as they determine the information outputs and/or physical effects on the state of the world as a result of service enactment.

-- A guard mediates access to the protected web service and enforces rules of corresponding access control policies. The guard must evaluate all requests, correctly evaluate policies, be incorruptible, and nonbypassable. Guard might be a standalone proxy or embedded as a wrapper of web service. Peculiarities of the mobile web service provisioning are in favor of the intermediate proxy implementation. This is a more general case and the functionality of guards can be integrated with resources.

-- A policy has rules that define which users may access the mobile web service. The goal is to isolate policy decision logic from resource and enforcement code. In the SBAC the policy is always an ontology or a set of ontologies that define semantic profiles of users, mobile web services, context, and policy rules of access.

-- A subject and object descriptors are well known patterns that provide access to the relevant attributes of subject and objects of access. A representation of descriptors for users, mobile web services, policies and context in the form of semantic annotations for mobile environments is reasonable because checking of attributes is independent from establishing them; there are different sources of attributes; different attributes are needed in different contexts; etc.

-- Context is a container for data that are relevant for access control decisions and enforcement. Different temporal and special characteristics like the time and location are traditionally considered as contextual information in mobile environments.

The components of the SBAC abstract architecture have to be deployed to the Mobile Host and middleware nodes depending on characteristics and requirements of use cases. There are several reasonable options of deployment of the SBAC components for protected mobile web service provisioning with unique characteristics and implications on the level of security and quality of mobile web service.

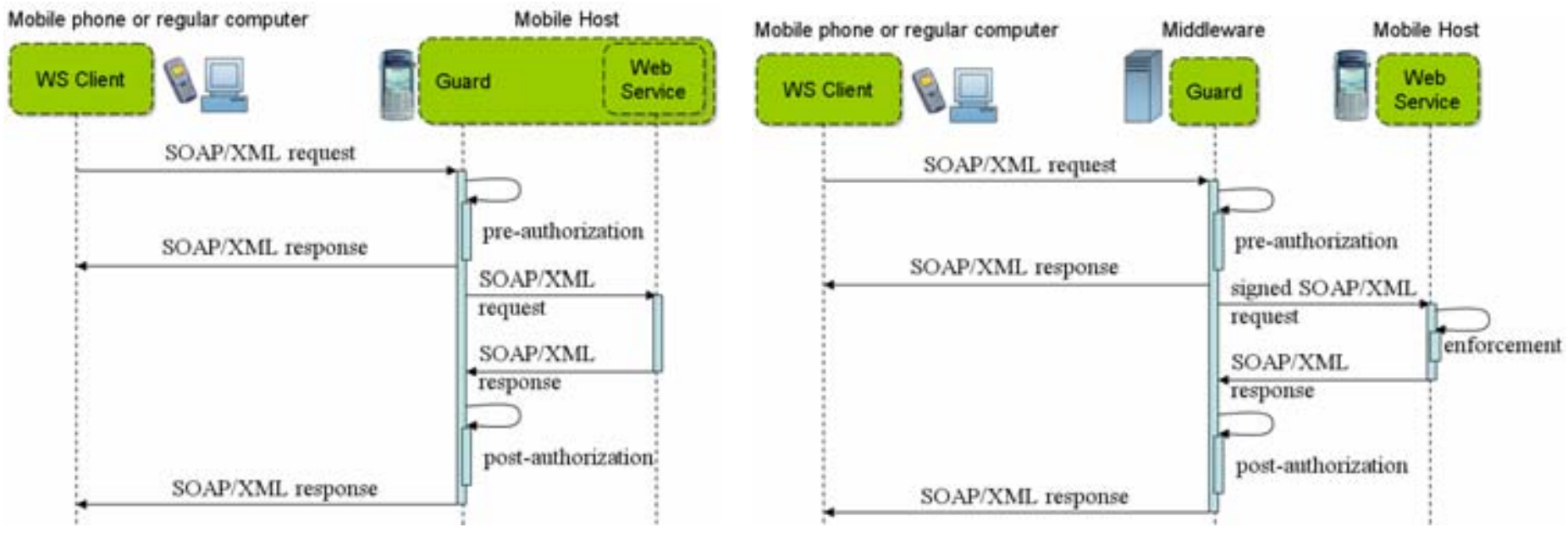

**Fig. 4.** Deployment options with 1. Embedded guard to the Mobile Host and 2. Middleware guard that is a proxy web service for the mobile web service

Figure 4 represents the first option with the embedded guard to the Mobile Host as a wrapper of the mobile web service. This is the most natural option for the pervasive mobile web service provisioning with the P2P communication between mobile clients and Mobile Hosts [10]. The clients access the service in the regular way. SOAP/XML messaging between the guard and service can be done using native RPC calls without delays of wireless or wired communication as other options have. One crucial advantage of this option is the opportunity to perform post-authorizations i.e. procedures of access control that must be performed after service enactment e.g. filtering of the content of response. This option supports the principle of end-to-end security in contrast to other options. However computational limitations of mobile phones demand light-weight functionality of the guard that prohibits use of complex semantics-based algorithms of reasoning for authorization decision making process.

Figure 4, with option 2, illustrates the deployment option where the guard is a middleware component and intermediate web service proxy that provides the same interface as the original mobile web service, decorates web service invocation with the SBAC policy enforcement mechanism, and delegates authorized requests to the mobile web service. The middleware guard is deployed in the Internet or mobile infrastructure. When the guard is in the Internet, clients are able to access it in a traditional way. Moreover Mobile Hosts receive a less number of requests or in other words only authorized requests, thus improving the scalability of the Mobile Host. The post-authorization is still possible. The middleware guard can aggregate and serve several Mobile Hosts and web services. Mobile-to-mobile requests experience delays of wireless communication twice when the guard is not an embedded but middleware component. There is a need to implement and deploy an enforcement component on the Mobile Host to validate signatures of the guard.

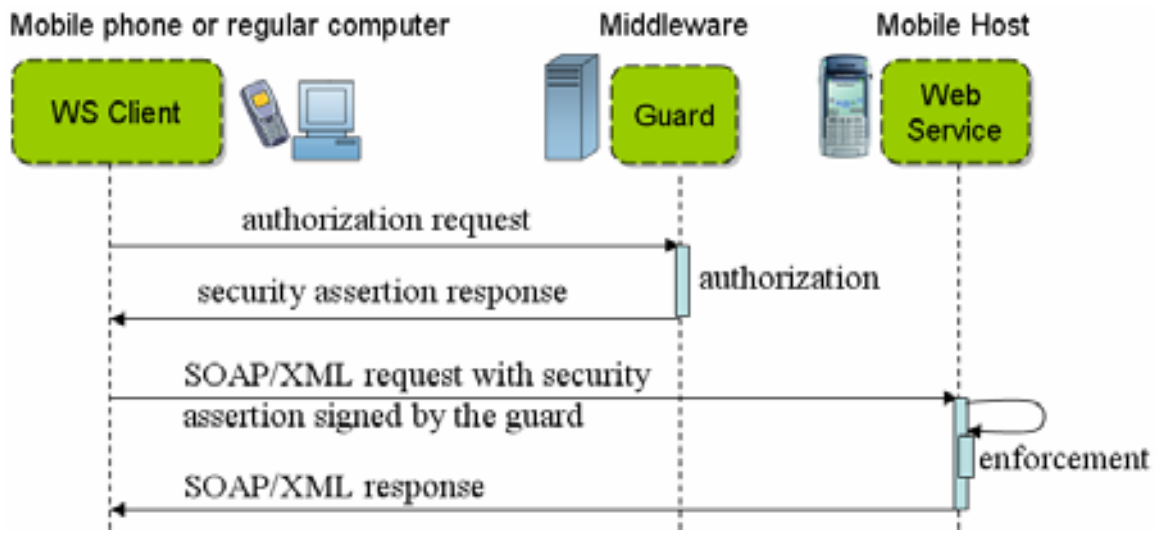

**Fig. 5.** Deployment option where guard is a third-party authorization authority.

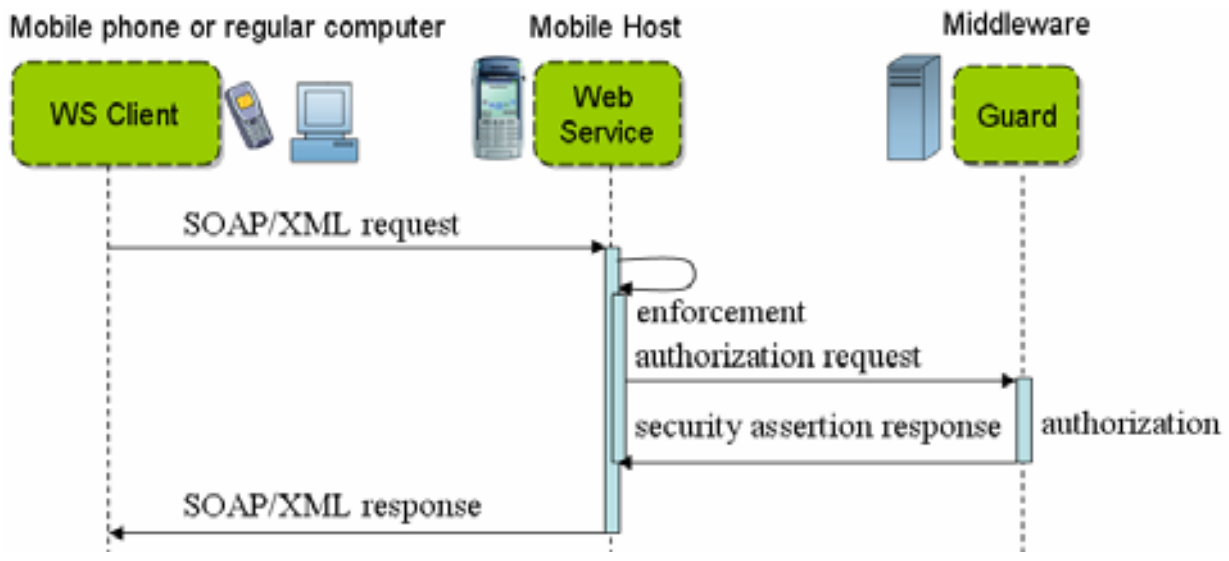

**Fig. 6.** Deployment option with the delegation of authorization to the middleware guard.

The deployment option, shown in figure 5, where the guard is a third-party authorization authority, creates additional inconveniences for clients to get authorization assertions prior to access protected mobile web services. Then enforcement components deployed on Mobile Hosts validate security assertions provided with requests in addition to verification of digital signatures in the previous option. The validation of security assertions is necessary to check that a security assertion corresponds to an operation for which a request is obtained when mobile web services provide more than one operation. Although this case might look too complex, however this is probably the most suitable option for the industrial, commercial or professional use of mobile web services when clients can get security tokens with long period of validity on the basis of their memberships in or subscriptions to different organizations, social networks, commercial services, etc. This option allows direct multiple requests to mobile web services using the same security token over time without overheads of the authorization decision making process for each request.

Delegation of authorization, when mobile web services initially receive all requests directly from clients and then generate requests for the access control decision to the middleware guard, is the last option we consider in this paper. Figure 6 depicts sequence diagram of communication between actors for this option. While such kind of deployment is possible, it has several significant shortcomings without clear advantages compared to above mentioned options. There are following needs: to embed the enforcement component for authorization messaging with all possible time overheads; to verify signatures of the guard; to process all requests from clients; etc. The only advantage is the shift of demanding functionality to the middleware guard.

## 5 Conclusion and Future Research Directions

This paper mainly discussed issues in providing secured communication and access control for the mobile web service provisioning. The paper first introduced the concept of mobile web service provisioning and then discussed the security breaches for the developed Mobile Host. It later presented the analysis of message-level security for the Mobile Host. The detailed performance analysis suggested that basic message-level security can be provided for the Mobile Host, even though not all the standards can be adapted to the mobile web service communication.

The paper later discussed the SBAC mechanism and adapting this mechanism in the provisioning of mobile web services, to achieve trusted and distributed management of access control for protecting mobile web services. Conducted analysis of deployment options reveals that they all are reasonable for realization and have different implications to security and QoS for the Mobile Host. However, future research in this domain, mainly addresses realizing the integrated security infrastructure. Further research on SBAC for MWS demands to assess all possible threats and attacks especially for the middleware guard. There is a need also to adopt SAML for the exchange of signed security assertions between WS clients, SBAC guards and Mobile Hosts. The adoption of SBAC for real-world application of MWS would also exemplify and align SBAC with practical concerns. Detailed performance analysis of the Mobile Host is again important, so that the extra load caused by the security mechanisms will not have serious impedances on the battery life of the devices and smart phone's basic purposes like making normal telephone calls.

**Acknowledgments.** This work is supported by the Research Cluster Ultra High-Speed Mobile Information and Communication (UMIC) at RWTH Aachen University (http://www.umic.rwth-aachen.de/) and partly funded by the grant from the Rector of the University of Jyväskylä, Finland.